\documentclass{emulateapj}
\usepackage{apjfonts}
\usepackage{natbib}
\usepackage{graphicx}
\bibpunct{(}{)}{;}{a}{}{,}

\def\linea{Fe~{\sc i}~$\lambda$6301.5~\AA}
\def\lineb{Fe~{\sc i}~$\lambda$6302.5~\AA}

\newcommand\linec{Fe~{\sc i}~$\lambda$15648~\AA}
\newcommand\lined{Fe~{\sc i}~$\lambda$15652~\AA}
\def\arcs{$^{\prime\prime}$}

\shorttitle{Quiet Sun Magnetic Fields}
\shortauthors{Dom\'inguez Cerde\~na, S\'anchez Almeida \& Kneer}

\begin{document}

\date{ }
\title{
  Quiet Sun magnetic fields from simultaneous inversions of
  visible and infrared spectropolarimetric observations
}

   \author{I. Dom\'\i nguez Cerde\~na}
   \affil{Instituto de Astrof\'\i sica de Canarias,
              E-38205 La Laguna, Spain}
   \email{itahiza@iac.es}

    \author{J. S\'anchez Almeida}
    \affil{Instituto de Astrof\'\i sica de Canarias,
              E-38205 La Laguna, Spain}
   \email{jos@iac.es}

    \and
    \author{F. Kneer}
    \affil{Institut f\"ur Astrophysik, Friedrich-Hund-Platz 1,
D-37077 G\"ottingen, Germany}
   \email{kneer@astro.physik.uni-goettingen.de}

\begin{abstract}

We study the quiet Sun magnetic fields using
spectropolarimetric observations of the infrared and visible
Fe {\sc i} lines at 6301.5, 6302.5, 15648 and 15653 \AA.
Magnetic field strengths and filling factors are inferred by the simultaneous
fit of the observed Stokes profiles under the MISMA hypothesis.
The observations cover an intra-network region at the
solar disk center.
We analyze 2280 Stokes profiles whose polarization signals are above noise in
the two spectral ranges, which correspond to 40\% of the field of view.
Most of these profiles can be reproduced only with a model atmosphere including 3
magnetic components with very different field strengths, which indicates
the co-existence of kG and sub-kG fields in our 1\farcs5 resolution elements.
We measure an unsigned magnetic flux density of 9.6~G considering the full field of
view. Half of the pixels present magnetic fields with
mixed polarities in the resolution element. The fraction of mixed polarities
increases as the polarization weakens.
We compute the probability density function of finding each magnetic field
strength. It has a significant contribution of kG field strengths, which
concentrates most of the observed magnetic flux and energy.
This kG contribution has a preferred magnetic polarity,
while the polarity of the weak fields is balanced.

\end{abstract}

\section{Introduction}\label{intro}

The quiet Sun can be defined as that part of the Sun far from magnetic
activity. If we exclude plages and network, the remaining quiet area
covers 90\% of the surface of the Sun, and remains almost unchanged
during the solar cycle \citep[e.g.,][]{har93}.
Few decades ago, this internetwork (IN) was thought to be
almost devoid of magnetic fields and consequently, its relevance for the
solar magnetism was not obvious \citep{liv75,smi75}.
However, with the present instrumentation, when the spatial resolution is
around 1$^{\prime\prime}$, the quiet Sun turns out to be full of magnetic fields
\citep{gro96,lin99,san00}, with an unsigned flux density measured with
techniques based on the Zeeman effect of the order of
10~G \citep[e.g.,][]{san00,lit02}.
The analysis of observations with a spatial resolution twice better yields
an unsigned flux twice larger \citep{dom03a}\footnote{However, see
\citet{lit04b}.}. There are reasons to believe
that these resolutions are not enough to completely resolve the magnetic
structures \citep[see][and references therein]{san03}.
Recent measurements based on the Hanle effect find an unsigned flux of, at
least, 60~G \citep{tru04,bom05}, which is
much higher than the magnetic flux observed with Zeeman techniques.
Considering the fraction of solar surface corresponding to the quiet Sun, the flux
measured in the IN is comparable to (or larger than) that coming from all active
regions even in the maximum of activity. Thus, the quiet Sun may play a significant
role in the physical processes governing the solar global magnetism
\citep[e.g.,][]{unn59,ste82,san02b,sch03b}.

The topology and the scale of variation of the quiet Sun
magnetic fields are not yet clear.
There is evidence for the existence of mixed polarities, at least when the
spatial resolution is 1$^{\prime\prime}$  \citep{san96,sig99,san00,lit02,kho02}.
The small-scale magnetic structures in the quiet Sun  seem to be smaller
than 50--100~km \citep[see][]{kho02,dom03b}.
This size assumes one single structure in the resolution element,
however, it is to be expected that
the dynamical influence of the granulation on the magnetic fields
produces a distribution of many smaller structures. This fragmentation shows up
in  MHD simulations of turbulent dynamos \citep{cat99a,emo01}, and 3D
magneto-convection \citep[e.g.,][]{ste02,vog03b,vog05}.
Recently, \citet{cam05} studied various mechanisms proposed to produce intense magnetic
concentrations and limited the radii of the magnetic flux tubes to less than 10~km,
which is approximately the magnetic diffusion length scale \citep{sch86}.
This scenario was already proposed by \citet{san96} with an atmosphere formed
by optically thin magnetic  structures 
with diverse properties
\citep[see also][]{san98c}. Such a scheme is called MISMA (Micro-Structured
Magnetized Atmosphere). An inversion procedure based on it has
been able to fit the whole variety of asymmetric Stokes
profiles\footnote{We use the standard Stokes parameters to define the polarization state:
$I$~for the intensity, $Q$ and $U$ for the two orthonormal states of lineal
polarization and $V$ for the circular polarization. The Stokes profiles of a
spectral line give the variation with wavelength
of these four parameters. If the spectral line is formed in a homogeneous atmosphere,
the Stokes profiles are expected to be symmetric or antisymmetric \citep[e.g.,][]{lan92}.}
produced by the quiet Sun magnetic fields \citep{san00,soc02}.

Observational studies based on infrared (IR) spectral lines~find
 { magnetic field strengths in the range below 1 kG
 \citep[{sub-kG;}][]{lin99,kho02}}, while works using
lines in the visible spectral range tend to find kG fields
\citep{gro96,sig99,san00,soc02,dom03a,dom03b,lit04b}.
Kilogauss fields are also inferred from the presence of
a considerable number of G-band bright points in the quiet Sun \citep{san04a,dew05}.
\citet{san00} put forward a solution for this apparent contradiction:
the co-existence in the resolution element of a distribution of field strengths
going from zero to kG.
They show how to reproduce the observed bias with the vertical gradient of
the field strength induced by the hydrostatic equilibrium in the atmosphere.
Another possibility with the same consequences is a horizontal gradient \citep{soc03}.
The presence of a range of field strengths from sub-kG to kG fields
bias the measurements based on IR Stokes profiles to sub-kG fields.
Such a particular behavior is due to the high sensitivity of the
IR lines to magnetic fields. The \linec\ \citep[used by][]{lin99,kho02} is
magnetically saturated\footnote{The Stokes $V$ signals no longer have a linear
  relationship with the magnetic field strength. The level of
  polarization stays constant while the splitting of the Stokes V lobes grows
  in proportion to the field strength.}
at 400~G while the visible pair \linea, \lineb\
\citep[e.g., used by][]{san00,dom03a} saturates at 1.5 kG.
The IR polarization signals coming from the strongest fields are spread in a
wide range of wavelengths and they produce little signal.
On the contrary, weak fields saturate the IR line and
a significant Stokes V signal is  built up in a narrow wavelength range.
Weak field and strong field components appear in the same wavelengths
in the visible lines so that their relative contribution is proportional
to their magnetic flux. The weak field component
can dominate over the strong component in the IR Stokes $V$ profiles, while
the strong fields dominate over the weak ones in the visible profiles.
Thus, an incomplete analysis of the IR profiles leads
to weak fields, whereas the same analysis of the visible lines shows strong fields.
In order to test this conjecture, \citet{san03c} presented simultaneous
spectropolarimetric observations in the visible and IR
ranges. An independent Milne-Eddington inversion of the visible and IR lines
suggests the co-existence of sub-kG and kG fields. However, a more sophisticated
and simultaneous inversion is desirable to provide further support to the conjecture
of \citet{san00} and \citet{soc03}.

This paper presents such simultaneous inversions under the MISMA approximation
of the visible and the infrared Stokes profiles used in \citet{san03c}.
As we discuss above, it seems to be appropriate for reproducing the complex
Stokes profiles produced by the quiet Sun.
Section~\ref{obs} summarizes the observations and the data
reduction.  Some properties of interest of the magnetograms obtained in
the two spectral ranges are shown in $\S$~\ref{magn}. The strategies for the
data analysis and the inversion are explained in $\S$~\ref{data}. We present the results in
$\S$~\ref{res}
{whose consistency is analyzed in $\S$~\ref{cons}.}
A final \S~\ref{con} discusses the main conclusions.


\section{Observations and data reduction}\label{obs}

We re-analyze the spectra used by \citet{san03c} whose properties are only sketched
in the original reference. For the sake of comprehensiveness, the data and reduction
are detailed here. The part of the reduction
in which we bring the visible and IR spectra to equal spatial resolution is new.

We use two telescopes for the simultaneous visible and IR
observation\footnote{Nowadays, the observation is possible using a single
telescope with TIP+POLIS \citep[VTT, Tenerife, see][]{kho05} and with SPINOR
\citep[DST, Sacramento Peak Observatory, see][]{soc05b}.}.
The observation was performed at the Spanish
Observatorio del Teide (Tenerife, Spain).
The IR spectra were gathered with the Tenerife Infrared
Polarimeter \citep[TIP,][]{col99,mar99}
operated at the German Vacuum Tower Telescope (VTT). Spectrograms of
the full Stokes vector ($I,Q,U,V$)
were obtained in the IR iron lines \linec\ (Land\'e factor $g_L$=3)
and \lined\ ($g_L$=1.53). The visible observations
were carried out with the French-Italian telescope THEMIS
(T\'el\'escope Heliographique pour l'\'Etude du Magn\'etisme et des
Instabilit\'es Solaires) using the spectropolarimetric mode MTR
\citep{mei85,ray93}. The lines analyzed are  the pair
\linea\ ($g_L$=1.66) and \lineb\ ($g_L$=2.5).
The atomic parameters of the four spectral lines are listed in
Table~\ref{tbatomic}.
In order to obtain spectra in a two-dimensional field of view (FOV), the solar
surface was scanned from East to West with a step size of 0\farcs5. The slit
width was set to 0\farcs5 in both telescopes and the scan consisted in 60
positions. To improve the signal to noise ratio, the integration time was set to
30~s per position.

\begin{deluxetable}{lcccc}
\tablecaption{Atomic parameters of the lines under study.}
\tablehead{\colhead{Wavelength [\AA]}& \colhead{$\chi$ [eV]\tablenotemark{a}} &
  \colhead{$Agf$\tablenotemark{b}}  & \colhead{Transition\tablenotemark{c}} & \colhead{$g_L$}}
\startdata
6301.499\tablenotemark{d} & 3.65 & 9.766$\cdot10^{-6}$ & $^5$P$_2$~~$^5$D$_2$& 1.67\\
6302.492\tablenotemark{d} & 3.69 & 2.630$\cdot10^{-6}$ & $^5$P$_1$~~$^5$D$_0$& 2.5\\
15648.515\tablenotemark{e} & 5.426 & 8.035$\cdot10^{-6}$&$^7$D$_1$~~$^7$D$_1$& 3\\
15652.874\tablenotemark{e} & 6.246 & 3.443$\cdot10^{-5}$&$^7$D$_5$~~$^7$D$_4$& 1.53\\
\enddata
\tablenotetext{a}{Excitation potential from \citet{moo66} and \citet{kur94}.}
\tablenotetext{b}{Abundance of Fe times statistical weight of the
lower level times oscillator strength from \citet[][visible lines]{gur81} and
from \citet[][IR lines]{bor03}.}
\tablenotetext{c}{Spectral terms from \citet{nav94}.}
\tablenotetext{d}{Wavelengths from \citet{hig62}.}
\tablenotetext{e}{Wavelengths from \citet{nav94}.}
\label{tbatomic}
\end{deluxetable}
\begin{figure*}[t]
\resizebox{\hsize}{!}{\plotone{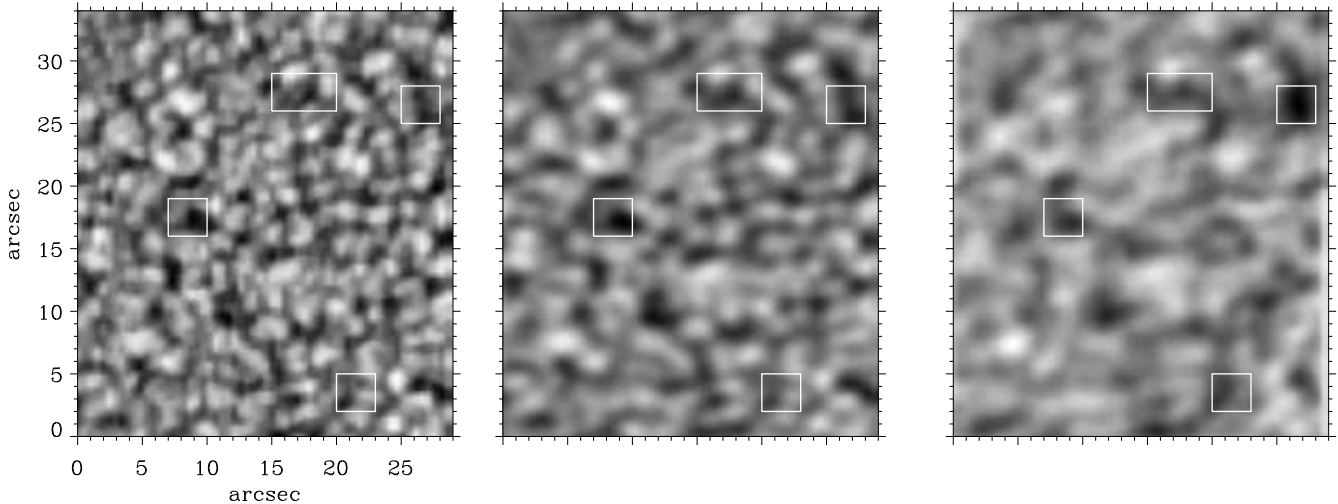}}
\caption{Intensity images in the continuum of the IR lines before
  (left) and after (middle) smearing, and continuum image of the visible lines
  (right). The rectangles point out structures that can be easily identified
  in both the IR and the visible images. Axes are in arcsec from the lower left corner.}
\label{img}
\end{figure*}

This work analyzes data obtained on August 10 2002 in a quiet Sun region close
to disk center ($\mu=0.94$). We choose a region without magnetic activity as
judged with the help of \ion{Ca}{2}~K and H$\alpha$ slit-jaw images from video cameras
at the VTT. A video link between both telescopes providing the slit-jaw images
from VTT at THEMIS was helpful to point both telescopes to the same target.
Since the absolute pointing of the telescopes was not accurate enough,
and the slit-jaw images of the granulation did not allow us to
recognize the region pointed by the two telescopes, we used another {\em differential}
method for accurate pointing.
First, both
telescopes were pointed to a characteristic small structure which
could easily be identified, e.g., a small pore.
The pointing of the VTT was moved to a quiet Sun region, and then
THEMIS pointing was displaced by the same distance.
As we explain below, the scans provided by the VTT and THEMIS are post-processed
for fine-tuning co-alignment. This final step removes residual pointing errors plus some additional effects (e.g., differential refraction between the visible and the IR).
The scan under study here has an offset between telescopes of only 1\arcs\
(see below). Since we scanned with a step size of 0\farcs5 and
with an exposure time of 30 s per step, the time lag between IR and
visible data of the same point corresponds to only 1 minute.

The data reduction of the two data sets was carried out in a similar way.
A flat-field correction was applied to the spectra.
In order to get the Stokes parameters, we demodulate the measurements with the calibration
matrix, and finally we correct for seeing-induced
crosstalk by combining the two beams of each
polarimeter.
The crosstalk produced by instrumental polarization (IP) of the telescopes
themselves was considered too. Since THEMIS is free of IP, the visible data do not need
correction. Due to the configuration of mirrors in the VTT, the IR data
are affected by an important crosstalk between the different Stokes
parameters. It was calibrated with a model Mueller matrix set using a linear polarizer
mounted at the entrance window of the telescope \citep[see][]{col03}.

\begin{deluxetable}{lcc}
\tablecaption{Convective blue shift of the spectral lines.}
\tablehead{\colhead{line} & \colhead{$v_{CB}$ \tablenotemark{a} [m\,s$^{-1}$]} &
  \colhead{$v_{CB}$ \tablenotemark{b} [m\,s$^{-1}$]}}
\startdata
\linea & +50 & -100\\
\lineb & -80 & -150\\
\linec & -370& -400\\
\lined & -430& -400\\
\enddata
\tablenotetext{a}{From synthesis in numerical simulations by \citep{asp00}.}
\tablenotetext{b}{From two-component inversion of many iron lines \citep[][]{bor02,bor03}.}
\label{tbCB}
\end{deluxetable}
\begin{figure*}[t]
\begin{center}
\epsscale{1.02}
\plotone{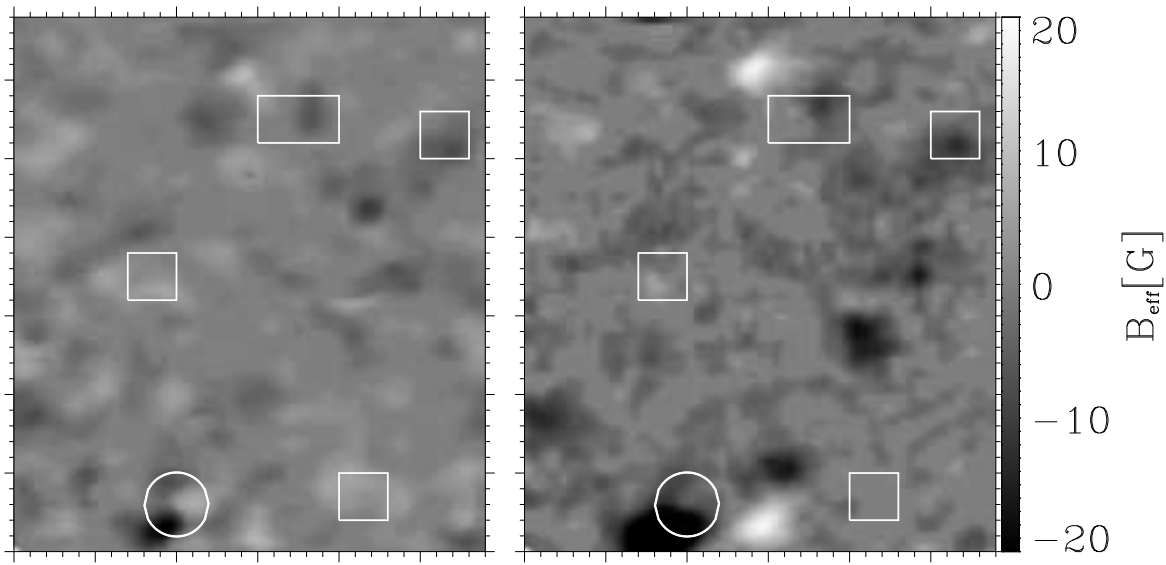}
\end{center}
\caption{Magnetograms from \linec\ (left) and \lineb\ (right). White and
  black distinguish the two polarities. Both magnetograms are
  scaled to flux densities $|B_\mathrm{eff}|<20$ G. The distance
  between small tickmarks is 1$^{\prime\prime}$. }
\label{virmag}
\end{figure*}

Since the purpose of this work is the simultaneous inversion of visible and IR lines,
one needs a consistent scale of wavelengths in both spectral ranges.
We carried out an absolute wavelength calibration so that the lines are unshifted in
this scale when the solar plasma is at rest. The central wavelength of
\linea\ and \lineb\
was considered to be the zero-crossing of the Stokes $V$ profiles corresponding
to the signals in the network points of our FOV. Such a reference
gives an accuracy of 200~m\,s$^{-1}$
{\citep{sol86}.
\citet{sol86b} find that the zero-crossing wavelength depends on the spectral resolution.
Using their simulations, we check that our  resolution of 21.9\,m\AA\ per pixel does not threaten the absolute accuracy given above.
}
The strongest polarization signals were chosen to trace the network
($V>0.02\,I_c$ with $I_c$ the intensity of the continuum near the lines).
Weaker signals tend to be red-shifted.
Since this method has not been tested for the IR lines, their absolute wavelengths need
another way of calibration. We computed the mean line profile from the data
and then corrected for the expected convective blue-shift.
We average the Stokes $I$ profile over the FOV.
The minima of the lines were fitted with fourth
order polynomials whose minima were taken as the central wavelengths of
the spectral lines. These wavelengths corrected for the convective blue-shifts
provide the reference for an atmosphere at rest.
Table~\ref{tbCB} presents the convective blue-shift
for \linec\ and \lined\ obtained from a two-component inversion
of many line profiles in the quiet Sun \citep{bor03} and from
line synthesis from numerical simulations of granulation
\citep[][private communication]{asp00}.
We cannot asses the accuracy of our absolute wavelength calibration based on them,
however, the fact that the two numerical simulations provide the same convective
blue-shift suggests the consistency of the approach.

One of the most important steps in the data reduction is the spatial
co-alignment of the visible and IR scans. The aim of this process is to
ensure that the scans are not only taken from the same region on the
Sun but also that each individual pixel of the visible and IR scans come from
the same point of the IN region.
The correct alignment must be guaranteed for a meaningful analysis of
the data, i.e., for the simultaneous interpretation of both data sets.
In order to align the scans we displace and re-scale them allowing for
five degrees of freedom. First, we correct for differences in the spatial scales
along the $X$ and $Y$ directions. We define $X$ as the direction of scanning and $Y$
as the direction along the slit.
The step width of the scans was slightly different in both telescopes
(less than 1\%), and the spatial scales along the slit were computed from
the focal length of the telescopes. Their (relative) agreement with respect to each other were checked using reference images of a sunspot.
Then, we found the shift in the $X$ and $Y$ directions by cross-correlation of the visible
and IR continuum images. It corresponds to the offset of
1$^{\prime\prime}$ mentioned above.
Finally both slits were not exactly oriented with the same angle with respect to the Sun.
There was a small rotation of 1$^{\circ}$ which we also determine by cross-correlation.
Once the correspondence between the visible and IR spatial coordinates was set,
we interpolate the visible spectra to obtain the spectra at the positions of
the pixels in the IR scan.

It is important to have the same spatial resolution in both data sets.
However, the angular
resolution of the  observations at the VTT was better than
at THEMIS due to the use of a correlation tracker
\citep{bal96}.  To solve this problem, the IR data were degraded to
the resolution of the visible data by convolution with a Gaussian.
{
The contrast of the IR continuum image depends on the angular resolution.
We set the width of the Gaussian, and so the angular resolution
of the IR map, by comparison with the synthetic images from numerical
simulations of magneto-convection \citep{vog03b,vog05}. Specifically, we tune the IR contrast so that the ratio between the visible and the IR contrasts agrees
with the ratio obtained from numerical simulations (0.48).
Such theoretical ratio was found to be almost independent
of the angular resolution and
in excellent agreement with the value measured by \citet{kho05} using a single
telescope to obtain simultaneous visible and IR data.
}
The central panel in Figure~\ref{img} shows the continuum image in the IR after smearing.
The contrast of the IR image is 0.9\%. The right panel of
Figure~\ref{img} contains the continuum intensity image of the visible
data (contrast 1.9\%). The squares point out regions easy to
identify in both images showing the goodness of the alignment.

After all data reduction, including smearing of the IR data and Fourier
filtering of the visible data, the level of noise in polarization for Stokes $V$
is (2-3)$\times10^{-4}I_c$ for the IR data and 5$\times10^{-4}I_c$ for the
visible data.
{ A more detailed description of the telescopes and the data reduction can be found in \citet{dom04}.
Descriptions of the instruments can be found in \citet{col99}
and \citet{sch02b} for TIP, and in \citet{lop00}
and \citet{bom02} for THEMIS.}


\section{Magnetograms}\label{magn}

Figure \ref{virmag} shows the magnetograms calculated from the Stokes~$V$ profiles
of \lineb\ and \linec.
The longitudinal magnetic flux density or $B_\mathrm{eff}$ was computed with the
magnetograph equation under the weak field approximation 
using the same method as \citet{dom03a}.
The magnetograms thus obtained are equivalent to the routine magnetograms
(e.g., 
that obtained with the MDI instrument on board of SOHO\footnote{See http://sohowww.nascom.nasa.gov/.}),
 if we exclude the velocity fields.
Figure~\ref{virmag} displays both magnetograms with the same scale of magnetic flux
densities. At first sight, both images are similar with the exception
that the signals found in visible are larger than the signals found in IR.
Indeed, this is the first important difference.
Since only the IN fields are of interest here, the network region pointed by a circle
in the lower left part of our FOV was discarded from the analysis by
rejecting pixels with flux densities larger than 20~G in the visible and
10~G in the IR. The noise levels turned out to be 2~G in the visible and 1~G in
the IR. {The method to estimate the noise is
detailed in \citet{dom03b}.}
The unsigned flux density obtained from pixels in the FOV with flux density
below the upper limit and above noise was,
\begin{equation}
\overline{|B_\mathrm{eff}|}=
\left\{\begin{array}{l}
 3.0~\textrm{G,  for \lineb}, \\
 1.5~\textrm{G,  for \linec}.
\end{array}\right.
\label{virm1}
\end{equation}
In both cases the pixels with signals cover 45\% of
the total FOV (the rest of the pixels were set to zero in the
calculation of $\overline{|B_\mathrm{eff}|}$). Under the same conditions of noise and spatial
resolution, the measurements in the visible detect more flux than in the IR.
This suggests that visible and IR lines trace different magnetic structures.
The difference of flux is conspicuous,
and one may be tempted to ascribe it to the approximations used for $B_\mathrm{eff}$.
However, more elaborated techniques show the same behavior \citep[ME inversions
  in][ and \S 5]{san03c,dom04}.

Some magnetic flux concentrations are very similar in
both magnetograms (squares in the upper part of Figure~\ref{virmag}) but
some others differ (see the other two squares). There are structures
having one polarity in the IR magnetogram and the opposite in the visible
one. This effect is evident even in the network area with large
polarization signals (see the circle in Fig.~\ref{virmag}). Such a
result can be considered as an additional
evidence of the different sensitivity to magnetic fields of the
visible and the IR lines.

The magnetic fields in the visible magnetogram (Fig.~\ref{virmag} right)
exhibit a preferred polarity while the IR magnetogram (Fig.~\ref{virmag} left)
looks well balanced. The signed flux was measured to be,
\begin{equation}
\overline{B_\mathrm{eff}}=
\left\{\begin{array}{l}
 -2.0~\textrm{G,  for \lineb}, \\
 -0.1~\textrm{G,  for \linec}.
\end{array}\right.
\label{virm2}
\end{equation}

This imbalanced flux is similar to that measured in other
visible observations \citep[e.g.,][]{lit02,dom03b}. Yet, \citet{kho02} found
almost no signed flux using the same IR lines (they observed 0.2~G).

All these results indicate that visible and IR lines may trace different
magnetic structures, as conjectured by \citet{san00,soc03}.

\section{Data Analysis}
\label{data}

In order to infer the magnetic field strengths from the observations, we
carried out an inversion of the Stokes profiles under the MISMA hypothesis.
MIcro-Structured Magnetized Atmospheres (MISMAs) were introduced for
the first time by \citet{san96}, with the objective
of reproducing the asymmetric Stokes profiles observed in most
solar structures. Unresolved optically thin micro-structures
naturally render asymmetric Stokes profiles.
The MISMA scenario turns out to be an appropriate scheme of analysis
since the observed quiet Sun Stokes profiles are very asymmetric,
and it has been shown to work with the asymmetries of \linea\ and
\lineb\ \citep{san00}.
In this section we give a description of the procedure followed to
invert the data.

\begin{figure*}[t]
\begin{center}
\includegraphics[height=19.2cm]{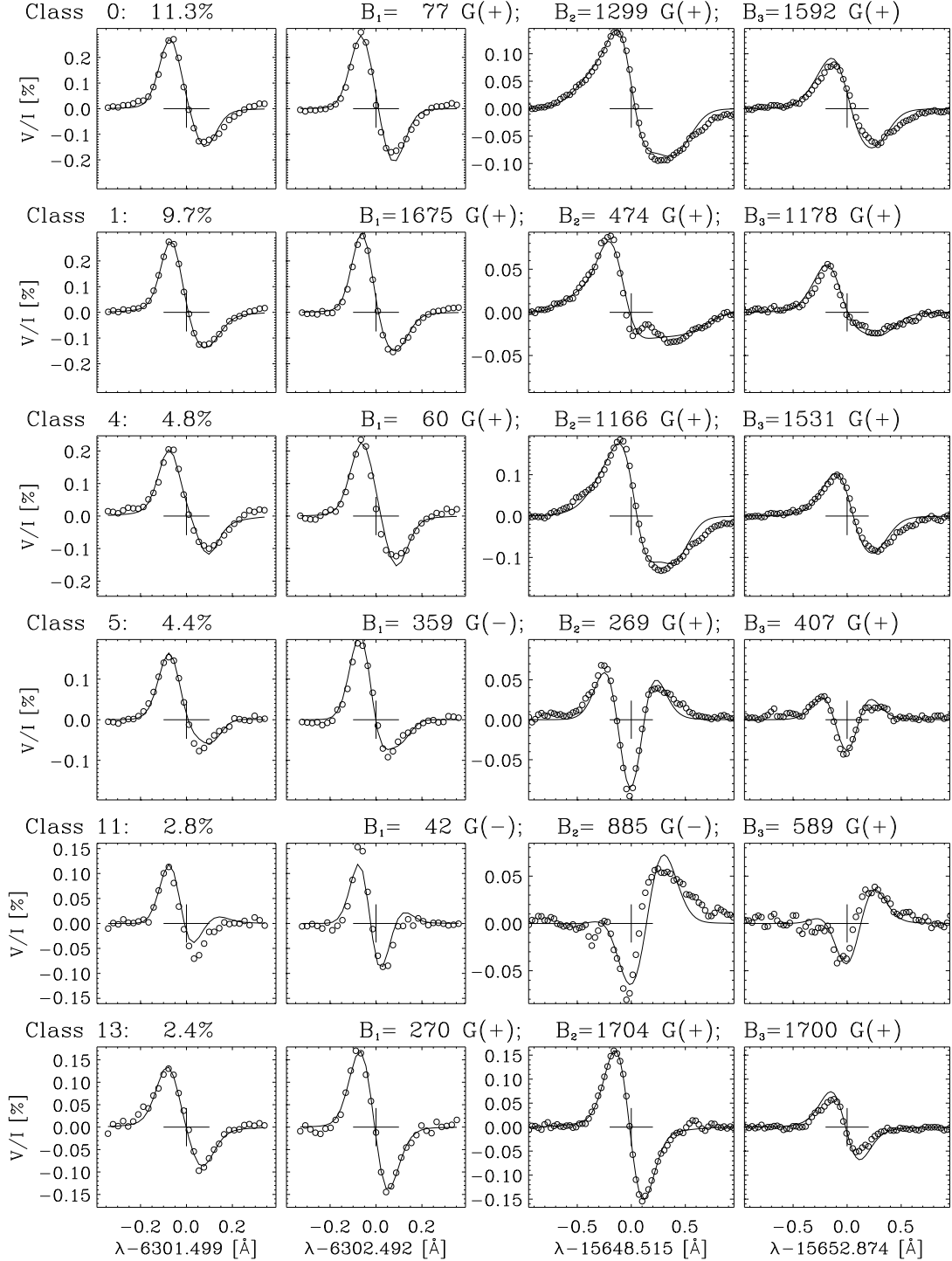}
\end{center}
\caption{Some representative classes of observed Stokes $V$ profiles
  showing the large variety of asymmetries produced by the quiet Sun magnetic
  fields. The circles represent the observed profiles and the solid lines the
  fits with MISMA model atmospheres. The header on top of each row displays the
  class number, the percentage of profiles belonging to the class,
  and the magnetic field of the three MISMA components at the base
  of the photosphere, sorted in order of decreasing mass. The polarity of the magnetic
  fields is indicated by plus and minus signs.}
\label{clas1}
\end{figure*}

\begin{figure*}[t]
\begin{center}
\includegraphics[height=19.2cm]{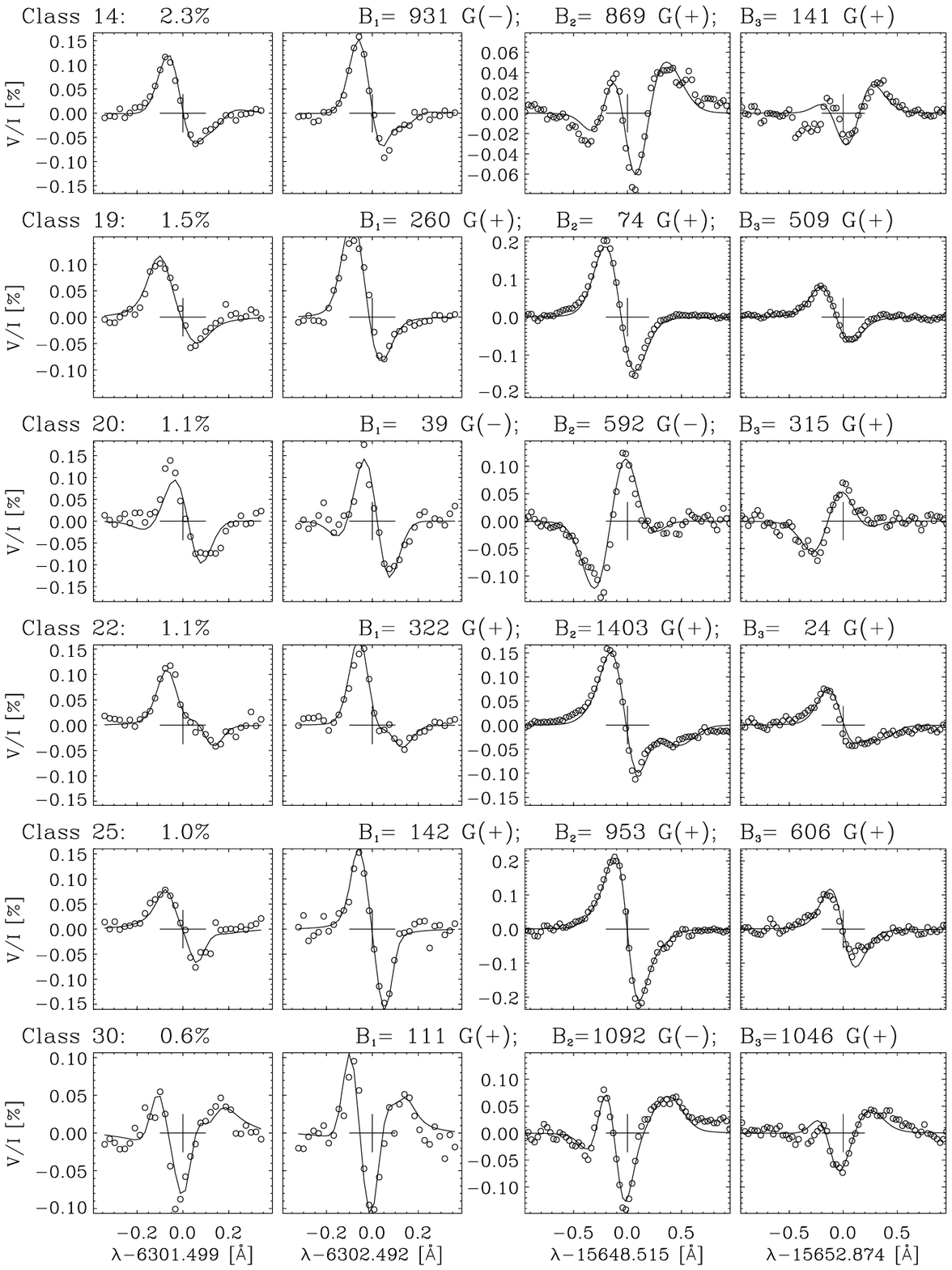}
\end{center}
\caption{Same as Figure~\ref{clas1} for 6 additional classes.}
\label{clas2}
\end{figure*}

\subsection{PCA classification}

We only analyze those Stokes $V$ profiles clearly
above noise. A threshold of three times the
noise level was chosen in order to separate clear polarization signals from
noisy profiles. Pixels with profiles whose peak polarization
is above the threshold in both the IR and the visible data cover
40\% of the FOV. This data set includes some 2280 pixels.

Following \citet[][\S~3.2]{san00} the Stokes $V$ profiles in the selected data set
were classified using a principal component analysis algorithm (PCA).
The aim is twofold. On the one hand, it reduces the number of
representative profiles and, on the other hand,
it provides low noise profiles to perform inversions used
as initialization for the inversion of individual Stokes profiles.
The classification is carried out simultaneously for the visible and the IR
Stokes $V$ profiles. The profiles of the four lines were scaled to the blue
lobe of the \lineb\ Stokes $V$ profile. The PCA method gives 38 classes.
All profiles with large polarization signals were included in one single
class to avoid contaminating the other classes with network profiles.
The Stokes $V$ threshold was chosen to be 1\%$I_c$ in the visible, and
this special class includes all profiles from the network patch in the
bottom left of our map (see Fig.~\ref{virmag}).
The large number of classes compared with those obtained for visible lines by
\citet[][10 classes]{san00} or for IR lines by \citet[][8 classes]{kho02} is due to the
combination of the different asymmetries found in the Stokes $V$ profiles of
the two pairs of lines.
Note that, in general, the classes found in this work do not correspond to
those obtained by \citet{san00} and by \citet{kho02}.

We represent each class with the average of all profiles corresponding
to the class. The individual profiles are not scaled to carry out the average.
We only multiply them by the sign of the blue lobe of \lineb.
Some of these classes are shown in Figures~\ref{clas1}~and~\ref{clas2} (the dots).
There is a clear improvement in the signal to noise ratio
(S/N, {defined as the maximum of the unsigned Stokes~$V$ profile divided by the noise})
from the individual profiles to those representing the class.
The individual profiles have a typical S/N between 3 and 10,
whereas the two first classes have a S/N of 100.
The IR profiles from classes~0 and 4 have tails
 extending far out from their extremes, as those predicted by \citet{soc03}.
These tails come from the contribution of kG field strengths.
Such profiles are similar to class~0
in \citet[][Fig.~1]{kho02}.
The first 6 classes include 42\% of the total number of profiles.
One conspicuous fact is that some classes (e.g., classes~11 and 20) show one polarity in
the visible and the opposite in the IR. Such a result was
expected from the magnetograms in Figure~\ref{virmag}. Approximately 20\%
of the profiles show the effect \citep[see also][]{san03c}.

\subsection{MISMA inversion}\label{Minv}

The MISMA inversion code is described by \citet{san97b}.
We will follow a strategy similar to \citet{san00}. Here we
summarize the main characteristics of the underlying model atmospheres:

\begin{figure*}[t]
\begin{center}{\includegraphics[height=4cm]{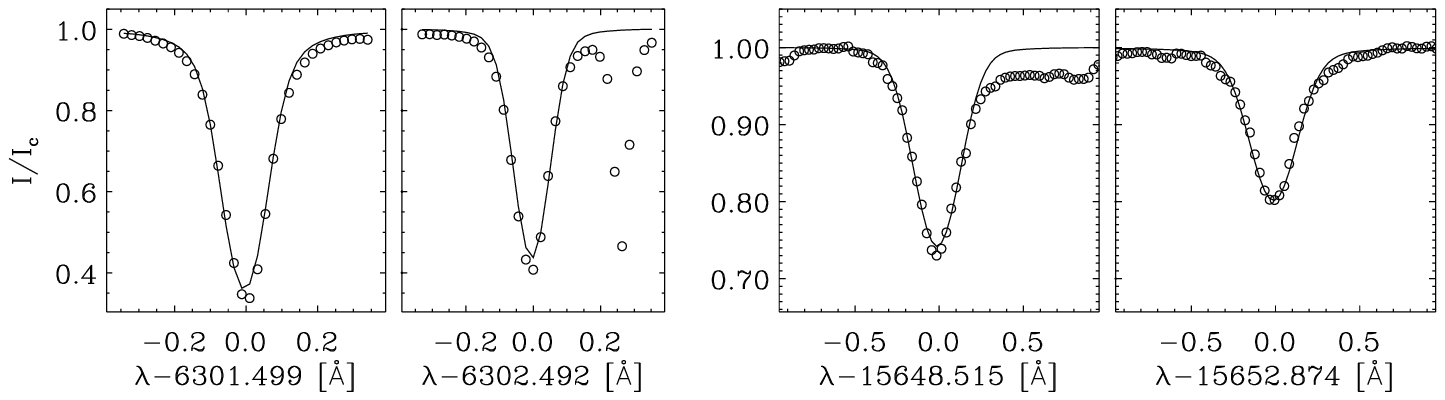}}\end{center}
\caption{Example of a fit of the Stokes $I$ profiles of the four lines.
 The circles represent the observed profiles, and
the solid lines are the fits. Note the two strong telluric blends
in the red wings of \lineb\ and \linec.}
\label{clasI}
\end{figure*}

\begin{figure}[t]
\resizebox{\hsize}{!}{\includegraphics{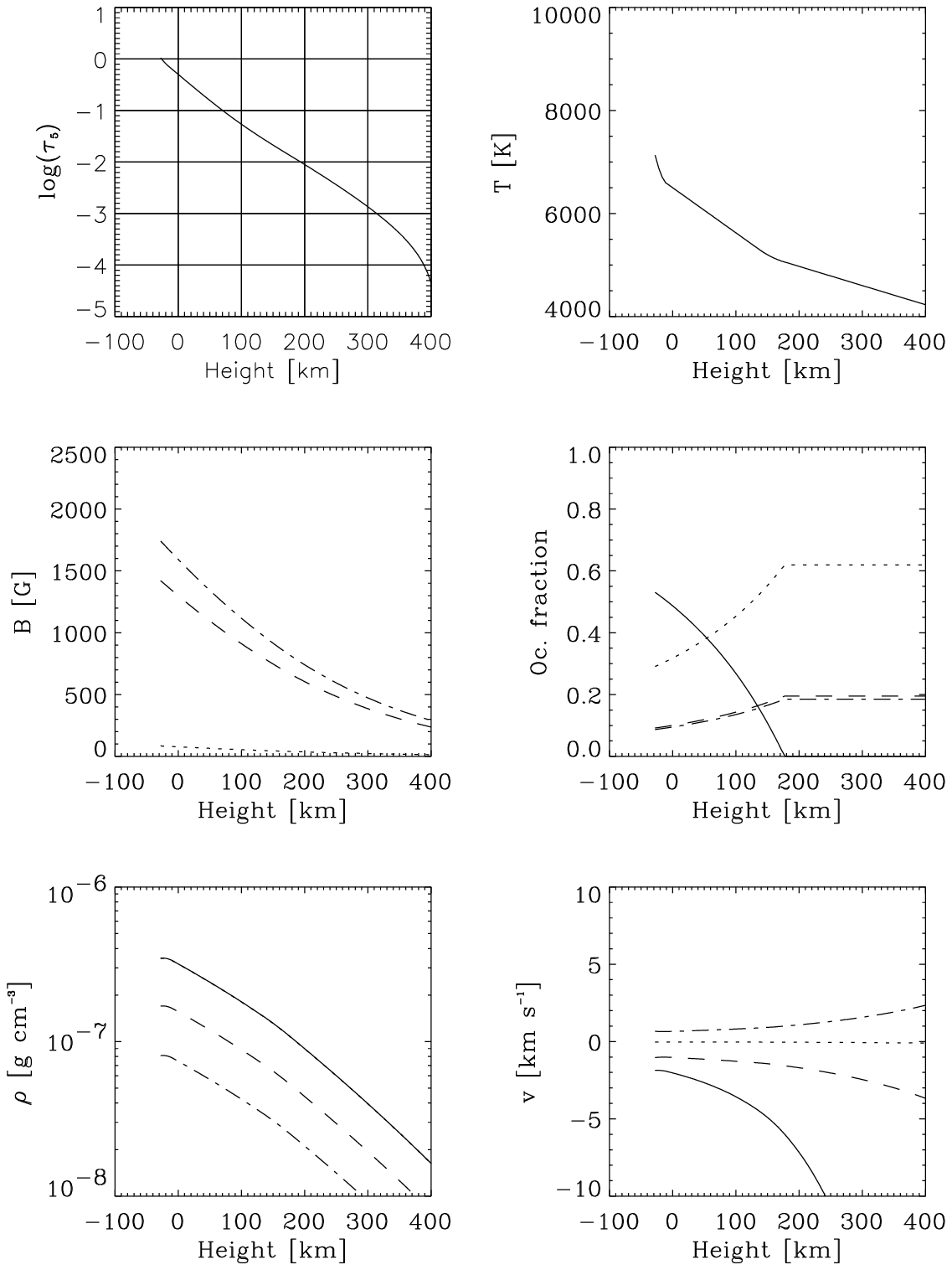}}
\caption{MISMA model from the inversion of class 0. The six panels show the
  variation with height of different physical parameters of interest. From left
  to right and top to bottom: the continuum optical depth at 5000~\AA,
  the temperature, the magnetic field strength, the occupation fraction, the density, and
  the velocity along the field lines. The parameters are displayed for the four
  components except the optical depth: the solid line corresponds to the
  non-magnetic component, and the dotted, the dashed, and the dot-dashed lines
  to the three magnetic components.
{Note that the spectral lines are formed between 100\,km and 300\,km,
  so that the properties of the atmosphere outside this range are not constrained
  by the observations and therefore are not reliable.}
  }
\label{mod}
\end{figure}

\begin{enumerate}

\item There are four different components, one non-magnetic and three magnetic.
Note that the inversion of visible lines performed by \citet{san00}
{
require two magnetic components. As \citet[][ \S 4.2]{san96} discuss,
it is impossible to reproduce even the mildest observed asymmetries
with only one component. Here
}
we need one additional component to characterize the weak magnetic field strength
revealed by the IR lines \citep{lin99,kho02,san03c}. The components are embedded in
a background producing unpolarized stray light.

\item The magnetic fields are vertical, along the line of sight. Since
the observed Stokes $Q$ and $U$ signals are rarely above the noise level,
it is reasonable to assume the inclination of the
magnetic field to be 0$^{\circ}$  or 180$^{\circ}$. The inversions
were performed only for Stokes~$I$ and Stokes~$V$.
This approach allows us to compute longitudinal magnetic fields, however,
the observations are compatible with inclined fields producing a linear polarization
 hidden by noise \citep[see][ \S~4.7]{san00}.
{The implications of this assumption are discussed in $\S$~\ref{incl}.}

\item The temperature of the components is forced to be the same. Since the
magnetic structures are optically thin the radiative exchange between
components is very efficient, washing out
temperature differences in seconds.

\item The four components are in lateral pressure balance.

\item Motions are only allowed along magnetic field lines.

\end{enumerate}

The temperatures are retrieved at four fixed heights in the
atmosphere. The parameters characterizing each component are set at the bottom  of
the model atmosphere. They are the magnetic field strength, the
occupation fraction, the bulk velocity, and the broadening parameter. The
stratification of these parameters is
obtained under the assumption of hydrostatic equilibrium
and horizontal pressure balance.
Three parameters account for the stray light, and another
one parametrizes both the solar macroturbulence and the
instrumental broadening due to limited spectral resolution.
{Despite the fact that the spectral resolution in the visible and the IR
differ by 30\,\%, one can use a single macroturbulence for
the two wavelength domains because the macroturbulence
inferred from the inversions is completely dominated by
true solar motions rather than instrumental smearing.
}
The {total} number of free parameters amounts to 25. They are
used to fit some 400 observables (each wavelength position of the
Stokes~$I$ and~$V$ profiles of the four lines).

An example of the fit to the intensity profiles is shown in Figure~\ref{clasI}.
The observed profiles (circles) exhibit line blends.
 These are a telluric line close to the red
wing of \lineb, a solar Fe {\sc i} line at 15647.3 \AA, and an unidentified
possibly telluric line in the red wing of \linec. To avoid the contamination
by these blends, the Stokes $I$ profiles at the wavelengths of the
blends are excluded from the fit.
There are also two OH bands of solar origin close to \lined, however,
these bands are only visible in cool solar regions such as sunspots.
The intensity profiles in Figure~\ref{clasI} are those from class~0.

All models have three magnetic components. In order to distinguish
between them, we denote them as first, second and third components.
The first component is that carrying the largest mass, as was defined by
\citet{san00}, and it is followed by the second and the third.
If the three components would have the same magnetic field strength,
the first would be the one producing the largest polarization signal.

Figure~\ref{mod} displays the model MISMA of class~0 (see the Stokes $V$
in Fig.~\ref{clas1}). The different physical parameters of the non-magnetic, the
first, the second, and the third magnetic components are plotted with solid, dotted,
dashed and dot-dashed lines, respectively.
The first magnetic component presents a weak magnetic field
strength ($\sim$100~G at the base of the photosphere),
while the other two have strong field strengths (larger than 1~kG).
The density of the first magnetic component is the same as for the
non-magnetic component (Fig.~\ref{mod} bottom-left panel) whereas
the other two components have much lower density. The reduction is forced by
the coupling between field strength and density produced by the
lateral pressure equilibrium. The higher the magnetic field the lower the gas pressure
and, for the same temperature, the lower the density.
%


\section{Results}\label{res}

Figures \ref{clas1} and \ref{clas2}  show the fits produced by
the MISMA inversion of 12 representative classes. All kinds of observed profiles
are properly reproduced with three magnetic components per resolution element.

From now on, all results correspond to a single height in the
atmosphere. We select the base of the quiet Sun photosphere,
defined as the height where the total pressure equals to
$1.3\times10^5$~dyn\,cm$^{-2}$.
Invoking lateral pressure balance among the individual model MISMAs, this
definition sets a single geometrical height for all the inversions,
and it corresponds to the height where the continuum optical
depth is one in the 1D quiet Sun model atmospheres \citep[e.g.,][]{mal86}.
The study is focused on two particularly important physical parameters,
namely, the magnetic field strength and the occupation fraction.
The latter corresponds to a volume filling factor.

The magnetic field strength inferred from the MISMA inversion
of each class is also given in Figures~\ref{clas1} and \ref{clas2}.
The magnetic field of the
components are denoted as B$_1$, B$_2$ and B$_3$, in order of decreasing mass.
Very often the inversions require one component with a sub-kG field
strength and two with strong kG fields. Such a behavior is typical of
the most abundant classes (e.g., classes 0, 1, 4). The \linec\ Stokes $V$
profiles of these classes are
similar to that expected from the synthesis by \citet{soc03}. The lobes
of the profiles have a splitting equivalent to that produced by a weak field,
however, they exhibit extended wings characteristic of strong fields.
Some other classes have all components with fields in the range
clearly below 1 kG (e.g., classes 5, 19, 20). In this case, there are no extended
wings in the profiles of \linec. Finally, note the presence of very asymmetric profiles.
The IR profiles tend to have a higher degree of asymmetry.
Some 28\% of the analyzed pixels present IR profiles with three lobes (e.g., class~5).
We found a fraction similar to that observed by \citet[][ 30\%]{kho02}.
There is also a small fraction of very asymmetric IR profiles with four lobes
(3\%, classes~14 and 30), while there are only a few
visible profiles with three lobes (3\%, e.g., class~30).

The model atmospheres obtained from the classes were used to initialize the inversion of
the individual Stokes $V$ profiles belonging to the class. From this moment on we refer only
to results obtained from the inversion of the 2280 individual profiles.

\subsection{Magnetic field strength and magnetic flux}\label{mags}

\begin{figure}[t]
\resizebox{\hsize}{!}{\includegraphics{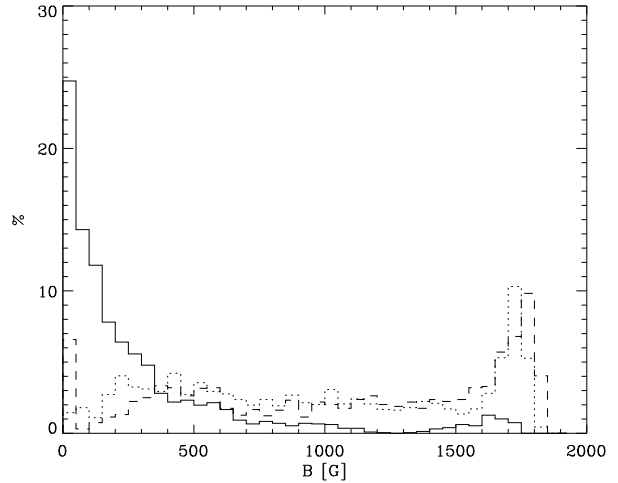}}
\caption{Histograms of the magnetic field strength at the base of the photosphere,
  determined from inversion of the individual Stokes $I$ and $V$ profiles.
  The first, the second and the third components
  are plotted with solid, dotted and dashed lines, respectively.}
\label{hmag}
\end{figure}

\begin{figure*}[t]
\begin{center}{\includegraphics[height=12.9cm]{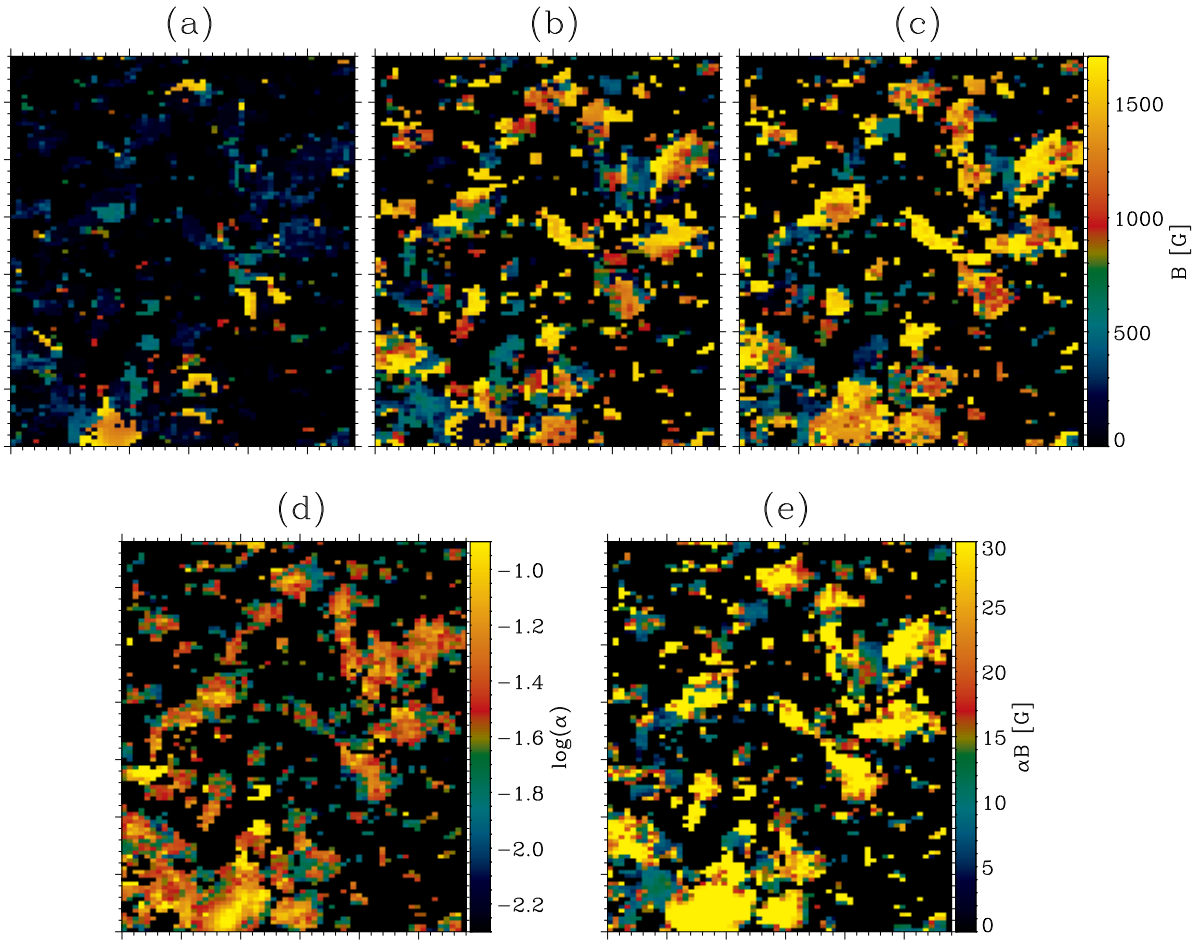}}\end{center}
\caption{The three upper panels (a,b,c) show maps of the
  magnetic field strength at the base of the photosphere of the first, the second
  and the third components, respectively. The bottom left panel (d) gives the
  occupation fraction (filling factor) of the three components all together
  in a logarithmic scale. The total unsigned magnetic flux is shown in the
  bottom right panel (e). It is saturated at 30~G. Black regions correspond
  to pixels with polarization signals too low to be inverted.
  The separation between minor tickmarks is one arcsec.}
\label{map}
\end{figure*}
\begin{figure}[!h]
\resizebox{\hsize}{!}{\includegraphics{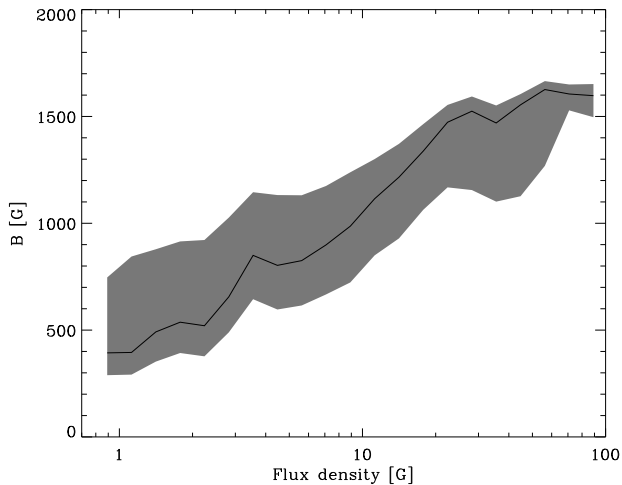}}
\caption{Mean magnetic field strength of all components as a function of the magnetic flux
  density. The shaded area gives the standard deviation above and below the mean value.}
\label{BF}
\end{figure}

The histograms of the distribution of magnetic field strength obtained from
the inversion of the individual profiles are displayed in Figure~\ref{hmag}. They
show the percentage of pixels within bins of 50~G for the
different components. The solid, the dashed,
and the dotted lines distinguish the first, the second, and the third magnetic
components, respectively. The results from the network patch were omitted (see \S~\ref{magn}).
The first component tends to have much weaker fields than the other two and it
has a shape similar to the distribution obtained by \citet{kho02} using only IR lines. The
mean field obtained for this component is 290~G. The
second and the third components show the same behavior as the histograms obtained
with visible lines by \citet{san00}. They have mean fields of 970 and 1100~G
respectively.
{
Despite the similarity between the magnetic field strengths of the second
and the third components, they are well separated by the inversion code.
There are large differences of Doppler shift between them; 85\,\% of the
models are separated by more than 500\,m\,s$^{-1}$ at the base of the photosphere,
and this difference increases substantially in the heights of
formation of the spectral lines. (See also $\S$~\ref{three}.)
}

The top panels of Figure~\ref{map} show the maps of field strength for
the three magnetic components. The magnetic field of the first, the second and the third
components are displayed from left to right. The network point (bottom left) is
almost the only feature with strong kG fields in the first component.
Most of the pixels show at least one component with strong fields. More
specifically, 92\% of the analyzed pixels possess at least one field larger
than 500~G while the fraction with field strengths above 1~kG is 74\%.

The map with the filling factor of the magnetic fields (the sum of occupation
fractions of the three components) is displayed in the bottom left panel of
Figure~\ref{map}. It goes from 10\% in network
points to 0.5\%, and the mean value is approximately 4\%. If we consider
the whole FOV (all values in the pixels not analyzed are set to zero), and we
exclude the network profiles, the measured magnetic fields fill only 1.5\% of
the IN region.

The bottom right panel of Figure~\ref{map} gives the map of the total magnetic
flux density computed as the product of the occupation fraction times the magnetic
field strength. Note that the flux density is given in G (or Mx\,cm$^{-2}$),
thus it can be directly compared with other measurements with different
pixel size or FOV. Most parts of the analyzed surface present
a flux density larger than 25~G (sum of the three components). Considering the whole
map, we measure a total flux density of 9.6~G which is similar to the value
obtained by \citet{soc02} or \citet{lit02} with better spatial resolution.

Figure~\ref{BF} shows the variation of the magnetic field strength of all components
with the magnetic flux density. The shaded area represents the standard deviation
above and below the mean value. A difference between the two standard deviations
reveals the asymmetry between the distribution of field strength above and below
the mean value.
There is a strong tendency for the field strengths
to weaken when the magnetic flux decreases. \citet{san00} obtained a
less pronounced dependence, but their lowest detectable flux density was 10~G
while in this work it decreases down to 1~G.


\begin{figure}[t]
\resizebox{\hsize}{!}{\includegraphics{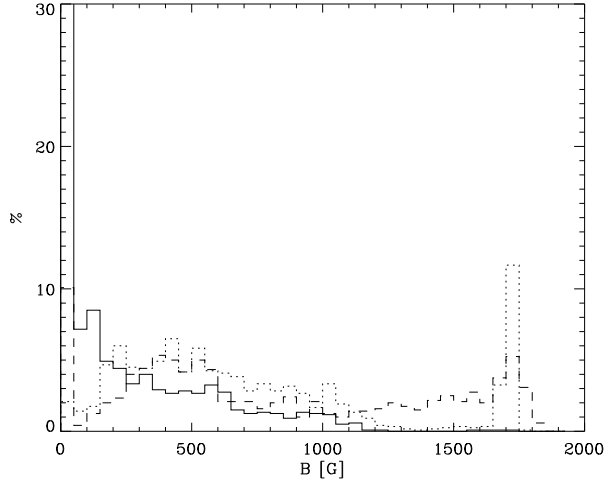}}
\caption{Same as Figure~\ref{hmag} but only for those pixels with mixed
  polarities in the resolution element.}
\label{hmag2}
\end{figure}

\begin{figure}[t]
\resizebox{\hsize}{!}{\includegraphics{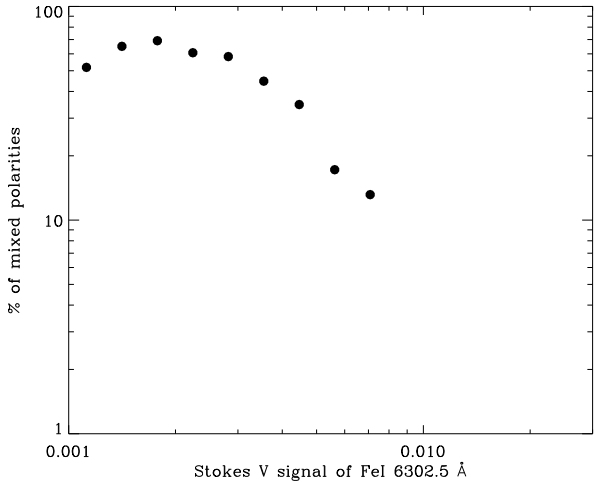}}
\caption{Percentage of pixels with mixed polarities versus the Stokes
  $V$ signal of \lineb.}
\label{mix}
\end{figure}
\subsection{Mixed polarities}\label{mixp}

50\% of the analyzed pixels can be fitted only with mixed polarities in the model MISMA.
Since we fit 40\% of the total pixels, this fraction corresponds to 20\% of the FOV.
This large amount of pixels includes all
profiles with three or four lobes in the IR and/or visible (30\%), and
also those profiles with a clear polarity in the visible Stokes profiles and the
opposite in the IR profiles (20\%).
We find that the fields of those pixels are weaker than the rest.
Figure~\ref{hmag2} shows the histograms of the field strength as in
Figure~\ref{hmag} but only for models including mixed polarities.
All components show fields weaker than in Figure~\ref{hmag}.
From the first, the second and the third magnetic components we retrieve
a mean field of 250, 700 and 880 ~G, respectively.


Figure~\ref{mix} gives the percentage of mixed polarities as a function of the
Stokes $V$ signal of \lineb. The weaker the polarization signal the larger
the fraction of mixed polarities.  This behavior was already noted by
\citet{san00} and \citet{soc02} but they obtained a maximum fraction of 25\%.
The fraction in Figure~\ref{mix} flattens to 65\% for
Stokes $V$ lower than 0.002.

\begin{deluxetable*}{lccccc}[!t]
\tablewidth{0pt}
\tablecaption{Filling factor, magnetic flux density and magnetic energy
  density of the PDF in
  Figure~\ref{pdf}.}
\tablehead{\colhead{Range}& \colhead{filling factor} & \colhead{flux density}  &
\colhead{energy density}  & \colhead{fraction of flux} & \colhead{fraction of energy}\\
\colhead{}& \colhead{[\%]}& \colhead{[G]}& \colhead{[erg\,cm$^{-3}$]}& \colhead{[\%]}& \colhead{[\%]}}
\startdata
total   &1.50 & 9.6 & 500 & 100 & 100\\
B<500~G &0.87 & 1.2 & 10 & 10 & 2\\
B>500~G &0.63 & 8.4 & 490 & 90 & 98\\
B>1000~G&0.42 & 7.0 & 450 & 75 & 90
\enddata
\label{tbpdf}
\end{deluxetable*}

\subsection{Distribution of field strengths}\label{spdf}

\begin{figure}[t]
\resizebox{\hsize}{!}{\includegraphics{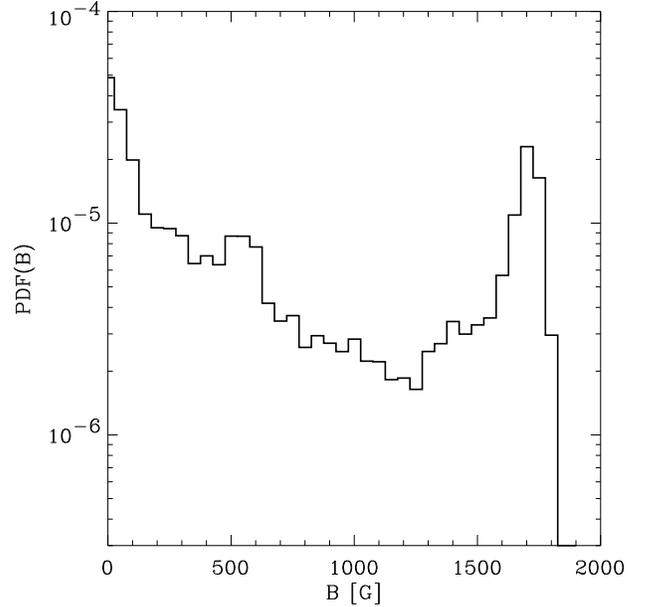}}
\caption{Probability density function of the magnetic field strength at the
  base of the photosphere as inferred from Zeeman signals.}
\label{pdf}
\end{figure}

We compute the probability density function (PDF) of the magnetic field
strength from the magnetic field and occupation fractions procured by the MISMA
inversions of the individual profiles.
The PDF gives the probability of finding a specific magnetic field
strength in the observed IN region. The range of magnetic fields from 0 to
2000~G was divided into bins of 50~G.
Each individual intrinsic field $B$ is assigned to one bin. The probability to
find the magnetic field of a single bin is the sum of the occupation fractions of
all pixels with $B$ belonging to this bin.
We normalize the PDF so that its integral provides the filling factor of
magnetic structures in the IN region assuming that the pixels without signals
have no magnetic field strength.

The PDF of the magnetic field strength condenses information on the quiet Sun
magnetism that  can be used for
comparisons with other observations and numerical simulations.
The first moment $\langle B\rangle$ is connected with the magnetic flux density and
the second moment provides the magnetic energy density ${\langle
  B^2\rangle}/{8\pi}$:

\begin{equation}
\langle B\rangle=\int_{0}^\infty B\ {\rm PDF(}B{\rm )}\ dB,
\end{equation}
\begin{equation}
\langle B^2\rangle=\int_{0}^\infty B^2\ {\rm PDF(}B{\rm )}\ dB.
\end{equation}

Figure~\ref{pdf} shows the PDF of our IN region. Once again, we have
avoided those pixels belonging to the small network patch in the FOV.
The shape of the PDF is to be
expected from the histogram of Figure~\ref{hmag}. Most of the filling
factor is concentrated in the weak fields, while most of the magnetic flux
comes from the strong fields. Table~\ref{tbpdf} summarizes the properties of
the PDF. It shows the filling factor, the flux density, the fraction of flux,
and the fraction of energy in four different ranges of magnetic field
strengths, i.e., the full range, the weak fields ($B<500$~G), the strong fields
($B>500$~G), and the kG fields ($B>1000$~G).
Three quarters of the total observed flux and 90\% of the magnetic energy is in
the kG fields, despite they represent only 30\% of the total filling factor.

There is an accumulation of field strength between 1600 and 1800 G.
It may be due to the combined effect of a magnetic intensification mechanism plus the
existence of an upper limit that the magnetic field strength cannot exceed.
This limit of 1800~G corresponds to a magnetic pressure equal to the
gas pressure of the non-magnetic plasma at the base of the photosphere.
The magnetic intensification mechanism would try to pile up magnetic fields
right before the upper limit.

\subsection{Signed flux}\label{sflux}

\begin{figure}[b]
\resizebox{\hsize}{!}{\includegraphics{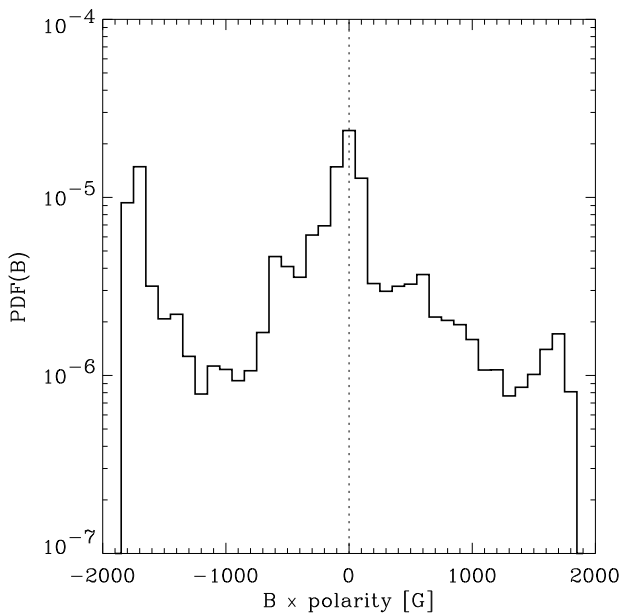}}
\caption{PDF of the magnetic field strength times the sign of the polarity of
  the flux. The vertical dotted line points out $B=0$~G.}
\label{pdf2}
\end{figure}

When we average over the FOV, the {\it signed} magnetic flux does not cancel completely.
The flux density, taking into account the direction of the magnetic
field, turns out to be $-3.5$~G. This net flux was
expected from the results in \S~\ref{magn} (equation~[\ref{virm2}]). \citet{lit02} already observed
the presence of an imbalanced flux in almost all the
observation he studied. He measured an imbalance (defined as the signed flux
divided by the unsigned flux) between $-0.20$ and $+0.48$,
and found a direct correlation between the imbalance of the IN fields and the
imbalance of the network fields around them.
Our imbalance is $-0.36$ and our network region is negative almost everywhere.
Figure~\ref{pdf2} shows the PDF of the signed magnetic field strength, i.e.,
the magnetic field strength times the sign of the vertical component of the
magnetic field. There is a large
contribution of strong fields in the negative flux part of the PDF, while the
positive part has a continuous fall off from weak to strong fields. The weak fields
are well balanced while the strong fields have a preference toward
negative polarity, i.e., the same polarity as the network patch. For fields
weaker than 500~G the imbalance is
only $-0.09$, and for field strengths stronger than 1~kG the imbalance is
$-0.5$. \citet{soc04} already observed the preference of the strong fields towards one
polarity, and Figure~\ref{pdf2} shows a clear observational
evidence of such an effect.
%

{
\section{Critical assessment of some assumptions
and results}\label{cons}

Some of the properties of the observational material
used in the work, and the assumptions of the inversion, are
not ideal but dictated by feasibility. In order to strengthen
the validity of the results, we devote this section to analyze
the assumptions and results that may be more critical.

\subsection{Time lag between IR and visible data sets}\label{time}

Our visible and IR spectra average the IN properties
in spatial scales of 1\farcs 5 arcsec and temporal
scales of 30~sec (\S~\ref{obs}).   The time lag between visible and
IR spectra would be of concern only if this spatio-temporal
average changes significantly during 1~min. Fortunately,
large changes of the average  properties are not expected
for a number of reasons. First, the integration time of the
individual measurements smears out the short time-scale
variations of the IN fields, and only variations in the
range between 30 sec and 1 min are of concern.
Second,  low spatial resolution observations of IN
magnetic fields show how many structures remain stable for
tens of minutes, and longer \citep[e.g.,][]{zha98,dom03b}. Obviously,
our time lag does not represent a problem for these structures.
In view of the results coming from high resolution observations
and numerical simulations of magneto-convection, this time stability
cannot not be  interpreted as the stability of individual
magnetic structures  but the stability of the average properties.
For example, a strong granular downdraft lives longer than the
granules and, during its lifetime, it continuously advects
magnetized plasma towards a specific point of the solar surface
\citep{ras03}. Observed with low resolution, one
would detect a magnetic patch that remains in place for a
long  period of time, despite the fact that it is formed
by many small-scale short-lived magnetic structures that
are  continuously advected and engulfed by the downdraft.
Finally, for the average magnetic properties to
present significant variations
during the time lag, the typical velocities of the magnetized
plasma would have to be unphysically high. Consider the horizontal
velocities required to sweep out during the time lag all
magnetic elements existing in a resolution element.
In this case the magnetic structures in the same resolution
element of the visible and IR maps will be completely different,
questioning any interpretation based on their simultaneity. The
required velocities would have to be of the order of the speed
needed to cross a resolution element during the time lag, i.e.,
18~km~s$^{-1}$. Such speed is too high to be common. It is supersonic,
and much larger than the observed vertical velocities.
Obviously, the lack of simultaneity becomes more problematic
as the spatio-temporal resolution of the observation improves.

From our point of view, the spatial and temporal
integration of the measurements represents a  problem more
serious than the time lag and the small spatial resolution differences
that may remain between the IR and the visible data sets,.
The polarization detected in each resolution element results
from adding up contributions from many magnetic
structures having very different properties.
Our inversions provide only a crude representation of
what may be happening in the real Sun. This problem, however,
is not specific of our work but affects all previous observational
works on the IN magnetic fields. In the case of the MISMA inversions
the problem is ameliorated because a sketchy model MISMA actually
represents many different underlying atmospheres, including
very complex ones. All those atmospheres having the same average
properties produce the same spectrum, and these average properties
are the ones that the inversions retrieve \citep[see the discussion in][\S~5]{san00}.

\subsection{The inclination of the magnetic fields}\label{incl}
The inversions were carried out under the assumption of
longitudinal magnetic fields, which minimizes
the number of free parameters and simplifies the
inversions. This assumption, however, biases the inferred
PDF($B$). Fortunately, the importance of the
bias can be estimated and  it turns out to be small.
If a magnetic field has an inclination $\gamma$ with respect
to our vertical line-of-sight, the Stokes $V$ spectrum remains
approximately as in the case of vertical field but scaled
with the factor $\cos\gamma$. This is a general property of
the radiative transfer equation, and is not restricted
to weak fields  on plane-parallel atmospheres
\citep[see][\S~3.1]{san99}. Consequently, if an observed
Stokes $V$ profile is reproduced with the wrong inclination,
all the parameters of the inversion would remain the same
except for the occupation fraction $\alpha(B)$. The occupation
fraction provides a global scaling factor for the Stokes
$V$ profiles, and so it absorbs the full bias.
Our  PDF($B$) is biased because it
is based on occupation fractions underestimated by a
factor $|\cos\gamma(B)|$,
\begin{equation}
{\rm PDF}(B)=N^{-1} \sum_{i=1}^{n(B)} |\cos\gamma_i(B)| \alpha_i(B),
	\label{pdfbias}
\end{equation}
with the subscript $i$ denoting one of the $n(B)$ pixels with a
field strength $B$, and the symbol $N$ standing for a normalization constant.
Obviously,  ${\rm PDF}(B)$ sets a lower limit to the true
${\rm PDF^*}(B)$,
\begin{equation}
{\rm PDF}^*(B)=N^{-1}\ \sum_{i=1}^{n(B)} \alpha_i(B)\ge {\rm PDF}(B),
\end{equation}
since $|\cos\gamma_i(B)|\leq 1$.
On the other hand, it is possible to set an upper limit by
estimating the maximum inclination allowed by the noise in
Stokes $Q$ and $U$. Following, \citet[][\S~4.7]{san00}, we
synthesize the linear  polarization produced by our model
atmospheres when the magnetic field of the three components
is inclined. For each model MISMA we determine the largest
inclination  $\gamma_u$ producing undetectable linear polarization, i.e.,
below the observational noise. If we use this inclination to
correct the occupation fractions,
one can define a new ${\rm PDF}_u(B)$ which represents an
upper limit to the true $PDF^*(B)$,
\begin{equation}
{\rm PDF}_u(B)=N^{-1} \sum_{i=1}^{n(B)}
{{|\cos\gamma_i(B)|}\over{|\cos\gamma_{ui}(B)|}} \alpha_i(B)
\geq{\rm PDF}^*(B) ,
\end{equation}
where the inequality follows from the fact that
$|\cos\gamma_i| /|\cos\gamma_{ui}| \ge 1$ for all $i$.
We find  that the unsigned flux corresponding to
${\rm PDF}_u(B)$ is only 25\% larger than the flux
of ${\rm PDF}(B)$, indicating that the assumption
of vertical fields have a very limited effect on the
PDF estimated in the paper.

\subsection{Need for three magnetic components}\label{three}

Due to the similarity between the magnetic field
strengths of the second and the third magnetic
components, one may wonder whether they are really
needed to reproduce the line shapes. Moreover, if they
are not needed, the existence of mixed polarities may be
an artifact since the signals of the two components
cancel our producing no observable residual.
As we stress in \S~\ref{Minv}, the three components are
needed. Here we elaborate on the reasons.
Reproducing the Stokes $V$ asymmetries of the visible
lines requires two magnetic components, even for the
mildest asymmetries characteristic of the network.
One single magnetic component produces Stokes $V$
area and the peak asymmetries of the same magnitude,
in contradiction with the observations
\citep[][p. 545]{san96}. Somehow
unexpectedly, \citet{san00}
find that these two components are enough to explain
not only the mildest asymmetries, but the whole range
of observations including extreme profiles with an
even number of lobes. These two components have similar
kG magnetic field strengths, but very different velocities
\citep[][Figs. 12 and 14]{san00}.
The large difference of velocities is demanded
by the observed Stokes $V$ asymmetries, and it is the
reason why the inversion code is able to separate
them. Here we analyze profiles similar to those
reproduced by \citet{san00} and,
consequently, a minimum of two magnetic components
is needed. In addition, we also want to reproduce
the IR lines which, according to the literature,
indicate the existence of weak fields in the quiet Sun.
A third weak magnetic field component was needed too.
This argument does not imply that all the observed
profiles necessarily require three components. It
implies that the typical cases need three components
(regular \linea\ and \lineb, with \linec\ and \lined\ showing
weak fields). It also implies that that typical Stokes $V$
profiles have enough information to constrain the
three components. Finally, it implies that if a single
MISMA scenario is going to be used to reproduce all the
asymmetries, it must contain three components.
}
\subsection{The inferred magnetic field strengths}\label{cons2}

One of the main conclusions of this work is the co-existence of kG and sub-kG
fields in resolution elements of 1\farcs5 as inferred from the simultaneous inversion of IR
and visible Stokes profiles.

The use of visible lines to measure kG magnetic field strength has been
criticized. The analysis of extremely noisy Stokes~$V$ profiles of \linea\ and
\lineb\ may induce a bias towards kG \citep[see][]{bel03}, and
the Stokes $V$ profiles from the visible lines are the major tracers of the kG
magnetic fields in our inversion.
However, the kG that we infer are not produced by such a bias for a number of
reasons.
The classes have much less noise than the individual profiles, and yet
the inversion yields as much kG of the individual profiles.
The classes that yield at least one kG magnetic component represent 75\% of
the total pixels. They include the  two main classes with a S/N
larger than 100. Another 10 classes have S/N larger than 30, which ensures
inversions without the bias pointed out by \citet{bel03}.

Another way to test the consistency of our results is to invert only the IR
lines to see whether it is possible to infer kG fields in the quiet Sun
without visible lines.
This inversion was performed in the same way as in \S~\ref{Minv} but only for
\linec\ and \lined\ with MISMA models including only 2 magnetic components
\citep[see more details in][]{dom04}.
\begin{figure}[t]
\resizebox{\hsize}{!}{\includegraphics{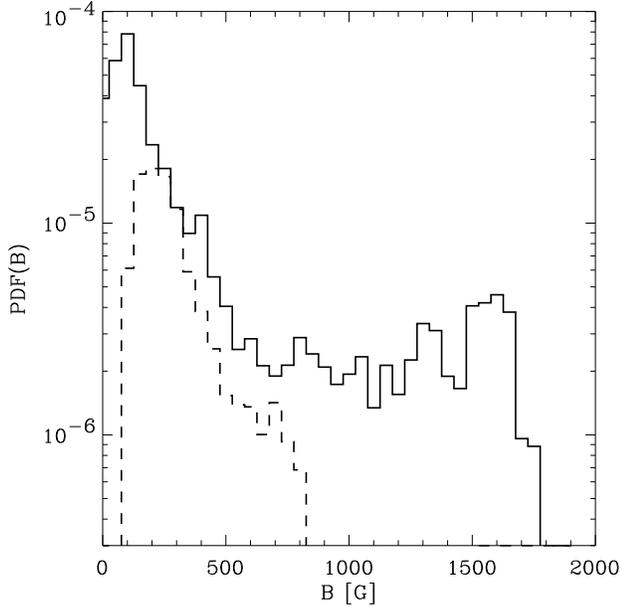}}
\caption{PDF of the magnetic field strength computed from the inversion of
the IR profiles alone. The solid line is the PDF inferred from the inversion under
the MISMA hypothesis, and the dashed line the PDF from the Milne-Eddington
inversion. }
\label{pdf3}
\end{figure}
The result of such analysis was a major magnetic component similar to the
first component in Figure~\ref{hmag} and a second with weak and strong
fields. A field stronger than 1~kG was found in 50\% of the inverted
pixels (those with the strongest IR signals).
The strong fields were retrieved from the extended wing of
the Stokes $V$ profiles. Figure~\ref{pdf3} gives the PDF
obtained from these inversions. The strong field contribution is not as large
as in Figure~\ref{pdf}, but it still carries most of the magnetic energy and flux.
The presence of fewer strong fields as compared to the full inversions
including visible lines is due to the incompleteness of the information given
by the IR lines, since most of the contributions from strong fields remains
hidden by noise.
The signed flux has a behavior similar to that of
the full inversion described in \S~\ref{sflux}. It is more
important for the fields stronger than 1~kG while the weak fields are
balanced. Figure~\ref{pdf3} also shows the PDF obtained from the
Milne-Eddington inversion of the IR lines \citep[the dashed line;
  see][]{san03c}. This inversion was performed with a single magnetic component
and shows only the weak field component.

\section{Discussion and Conclusions}\label{con}

A new  observational technique was used to study the
quiet Sun magnetism. Simultaneous visible (\linea\ and \lineb)
and infrared (\linec\ and \lined) polarimetric
observations were analyzed under the assumption of Micro-Structured
Magnetized Atmosphere (MISMA).
We generate model MISMAs able to reproduce all kinds of
asymmetries presented in the observed Stokes $V$ profiles, even when there are
clear differences in the asymmetries of the visible and the IR Stokes profiles.
The model atmospheres have three magnetic components.

  The simultaneous inversion of the visible and IR pairs reveals
     the co-existence of kG and sub-kG fields in the low photosphere.
     We have observed this combination of weak and strong fields in 74\% of
     the pixels under study, which means 30\% of the pixels of the FOV (see
     \S~\ref{mags}). This was
     already observed by \citet{san03c} from the separate inversion of the IR
     and visible lines, however, this work presents the first evidence
     of the co-existence of kG and sub-kG fields from a
     simultaneous inversion of lines in the two spectral ranges. From the
     three magnetic components, the one having most of the mass includes most of the weak
     fields. The two other components tend to show strong kG fields.
{The preference of the first component to have the weakest
 fields is due to the coupling between field strength and plasma density.
 The weaker the field the larger the density so that the component of weakest field
 tends to have the largest mass.}
     The histogram of field strengths of these minor components (Fig.~\ref{hmag})
     looks similar to that obtained by \citet{san00}, while the first component has
     a distribution similar to that obtained with the analysis of IR lines
     by \citet{kho02}. This supports the hypothesis of \citet{san00} and
     \citet{soc02,soc03}: the study of the quiet Sun magnetism with only
     visible lines or only IR lines biases the resulting field strengths;
     the IN fields must be studied combining Stokes profiles
     from both spectral ranges.

{
    The inferred magnetic fields occupy 1.5\,\% of the total FOV.
Other observations with better spatial resolution
\citep[1$^{\prime\prime}$;][]{soc02,kho02} yield a total filling factor of 1\,\%.
We obtain a larger value thanks to the combined use of the information
from visible and IR lines, while all previous works used lines from a single
spectral range.
The filling factor inferred for the same observations used in this paper by \citet{san03c}
was only 0.85\%. The difference is due to the different model atmospheres used to reproduce
the observations.
While here we use a realistic multicomponent model, \citet{san03c} carried out
Milne-Eddington inversions with a single magnetic component. Thus, the
filling factors in pixels with mixed polarities
(see \S~\ref{mixp}) are underestimated.}

  We have measured an average flux density over the FOV of 9.6~G
     (those pixels without polarization signals contribute to this average
     with zero field strength; \S~\ref{mags}). This flux is similar to that obtained by
     analysis of visible lines with  1$^{\prime\prime}$
     spatial resolution \citep[see][]{soc02,lit02}  while it is two times
     larger than the magnetic flux derived from IR lines \citep{kho02}.
     Albeit our observations have lower resolution (1\farcs5),
     the use of visible and IR lines reveals more magnetic structures
     than the use of lines from a single spectral range.
{The flux density is also larger than the magnetic flux obtained in
\S~\ref{magn} using the magnetograph equation applied to a single line.
A combination of factors explain such bias. First, we obtained magnetic
field strengths larger than the saturation limit of the lines,
thus the circular polarization signals are no longer proportional to the
magnetic flux and equation~(\ref{virm1}) is a lower limit of the real flux.
Second, the presence of mixed polarities (see \S~\ref{mixp})
produces cancellation of polarization signals and the magnetic flux is
underestimated.
Third, the lines weaken in strong magnetic concentrations,
leading to a well known diminishing of polarization signals \citep[e.g.,][]{har69}.
Finally, the visible and IR signals often trace different concentrations
coexisting in the resolution element. Then, combined inversions allow us to
detect more flux than the inversion of one of the spectral ranges.
}

  The inversion indicates the presence of mixed polarities in at
     least 20\% of the FOV (\S~\ref{mixp}). This is the largest fraction of
     mixed polarities obtained from direct measurements. The fraction of
     pixels with mixed magnetic polarities grows as the polarization
     signals weaken (Fig.~\ref{mix}). More than half of the profiles with
     Stokes $V$ signal lower than 0.2\% of the continuum intensity have  mixed polarities.

  We have estimated the probability of finding a given magnetic field strength
     in the quiet Sun, i.e., the probability density function of the magnetic field
     strength (PDF; \S~\ref{spdf}). It is largest for the weakest fields, however,
     an extended kG tail goes up to a limit around 1800~G.
     Despite the fact that the kG fields have only
     30\% of the total filling factor, they carry
     75\% of the total flux and 90\% of the total magnetic energy.

   There is a clear imbalance of the magnetic flux towards one
     polarity. We measure a signed flux density of $-3.5$~G which is mostly due to the
     strong kG fields  (\S~\ref{sflux}). The signed PDF (Fig.~\ref{pdf2}) shows
     how the weak fields are well balanced whereas the strong fields
     have a clear preference for the negative polarity. This
     imbalanced flux has the same sign as the surrounding network, as it was
     already observed by \citet{lit02}.
{This imbalance is qualitatively similar to those imposed in some
MHD simulations of granular convection \citep[see][]{vog03b,vog03}. These
simulations have an initial unipolar field. The convection reprocesses
this field to yield a continuous PDF with a clear imbalance in the strongest fields.}

  The MISMA inversion of the IR profiles alone also gives rise to a PDF with
     extended kG tail (\S~\ref{cons2}). Such tail does not appear in Milne-Eddington
     inversions of the same profiles, which tend to select the component of
     weakest field in the resolution element.

The magnetic fields studied in this paper fill only 1.5\% of the total
FOV. The unsigned flux density that we measure is only a lower limit of
the real flux. Larger fluxes are obtained in
observations  with better spatial resolution \citep{dom03a,dom03b,san03d} and, furthermore,
measurements based on the Hanle effect are compatible with a flux density of
60~G or more \citep{san03,tru04,bom05}. A large fraction of
the magnetic structures remains undetected in Zeeman based observations.
Based on our biased observations, we have produced a PDF with the
distribution of magnetic field strengths. Obviously, this is not the real solar PDF.
A natural step forward would be working out an unbiased
PDF with the help of different measurements based on the Zeeman effect,
the Hanle effect, and also MHD simulations. This work of synthesis
is carried out in a separate paper \citep[][]{dom06}.
{
Despite these and other efforts, we are still far from having a
complete picture of the quiet Sun magnetism.
There is a reduced number of observations and each one is gathered with different instruments
under different conditions.
Thus, we need a survey of polarimetric data across many IN regions of the Sun, if
we want to study the quiet Sun with a systematic and global point of view.
Such data would allow us to start exploring the connection
of the IN with active regions, the chromosphere and corona and,
eventually, to follow the evolution of the IN
with the solar cycle. Observations with better spatial and
temporal resolution are mandatory. They are coming up thanks
to the advent of new generation ground based and space-borne solar
telescope \citep{kei03}. These improvements,
together with the development of more realistic numerical
simulations, guarantee a rapid evolution of the field.
}

\acknowledgments

Thanks are due to the support astronomers and operators of THEMIS and
VTT for help during the observation.
The VTT is operated by the Kiepenheuer-Institut f\"ur
Sonnenphysik, Freiburg,
and the French-Italian telescope THEMIS is operated by CNRS-CNR, both
at the Spanish Observatorio del Teide of the
Instituto de Astrof\'\i sica de Canarias.
The work has been partly funded by the Spanish Ministry of Science
and Technology, project AYA2004-05792 and by the Deutsche Forschungsgemeinschaft through grant 418 SPA-112/14/01.



\end{document}